\documentclass[twocolumn,times, tighten]{aastex631}
\usepackage{scalerel,graphicx,xparse}
\usepackage{latexsym}
\usepackage{bm}
\usepackage[english]{babel}
\usepackage{booktabs}
\usepackage{amsmath}
\usepackage[Symbol]{upgreek}	
\usepackage{natbib}	
\usepackage{longtable}
\usepackage{array}
\usepackage{blindtext}

\newlength{\txw}
\newlength{\txh}

\newcommand{\fesc}   {\ensuremath{f_{\rm esc}}}

\newcommand{\HST}    {\emph{HST}}

\newcommand{\mAB}    {\ensuremath{m_{\text{\tiny AB}}}}

\newcommand{\zmean}  {\ensuremath{\langle z\rangle}}

\newcommand{\lyc}    {\textnormal{\fontsize{6}{6} \textsc{L}\text{y}\textsc{C}}}
\newcommand{\uvc}    {\textnormal{\tiny \text{UVC}}}
\newcommand{\SExtractor}{\textsc{SExtractor}}
\newcommand{\PreserveBackslash}[1]{\let\temp=\\#1\let\\=\temp}
\newcolumntype{C}[1]{>{\PreserveBackslash\centering}p{#1}}
\newcolumntype{R}[1]{>{\PreserveBackslash\raggedleft}p{#1}}
\newcolumntype{L}[1]{>{\PreserveBackslash\raggedright}p{#1}}
\NewDocumentCommand\cloud{}{\scalerel*[.9em]{\includegraphics{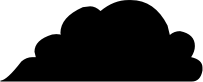}}{X}}

\defcitealias{Smith2020}{S20}
\defcitealias{Smith2018}{S18}
\defcitealias{Planck2020}{Planck Collaboration (2020)}

\begin{document}
\title{Lyman Continuum Emission from AGN at 2.3$\lesssim$z$\lesssim$3.7 in the UVCANDELS Fields}
\correspondingauthor{Brent M. Smith}\author[0000-0002-0648-1699]{Brent M. Smith}\affiliation{School of Earth \& Space Exploration, Arizona State University, Tempe, AZ 85287-1404, USA}\email{bsmith18@asu.edu}

\author[0000-0001-8156-6281]{Rogier A. Windhorst}\affiliation{School of Earth \& Space Exploration, Arizona State University, Tempe, AZ 85287-1404, USA}

\author[0000-0002-7064-5424]{Harry Teplitz}\affiliation{Euclid NASA Science Center at IPAC, California Institute of Technology, Pasadena, CA 91125, USA}

\author[0000-0001-8587-218X]{Matthew Hayes}
\affiliation{Stockholm University, Department of Astronomy and Oskar Klein Centre for Cosmoparticle Physics, AlbaNova University Centre, SE-10691, Stockholm, Sweden}

\author[0000-0002-9946-4731]{Marc Rafelski}
\affiliation{Space Telescope Science Institute, Baltimore, MD 21218, USA}
\affiliation{Department of Physics and Astronomy, Johns Hopkins University, Baltimore, MD 21218, USA}

\author[0000-0001-5414-5131]{Mark Dickinson}
\affiliation{NSF's NOIRLab, Tucson, AZ 85719, USA}

\author[0000-0001-7166-6035]{Vihang Mehta}
\affiliation{IPAC, Mail Code 314-6, California Institute of Technology, 1200 E. California Blvd., Pasadena CA, 91125, USA}

\author[0000-0001-6145-5090]{Nimish P. Hathi}
\affiliation{Space Telescope Science Institute, Baltimore, MD 21218, USA}

\author[0000-0001-6529-8416]{John MacKenty}
\affiliation{Space Telescope Science Institute, Baltimore, MD 21218, USA}

\author[0000-0003-3466-035X]{L. Y. Aaron Yung}
\affiliation{Astrophysics Science Division, NASA Goddard Space Flight Center, Greenbelt, MD 20771, USA}

\author[0000-0002-6610-2048]{Anton M. Koekemoer}\affiliation{Space Telescope Science Institute, Baltimore, MD 21218, USA}

\author[0000-0002-2390-0584]{Emmaris Soto}
\affiliation{Computational Physics, Inc., Springfield, VA 22151, USA}

\author[0000-0003-1949-7638]{Christopher J. Conselice}
\affiliation{Jodrell Bank Centre for Astrophysics, University of Manchester, Oxford Road, Manchester M13 9PL, UK}

\author[0000-0003-1581-7825]{Ray A. Lucas}
\affiliation{Space Telescope Science Institute, Baltimore, MD 21218, USA}

\author[0000-0002-9373-3865]{Xin Wang}
\affiliation{School of Astronomy and Space Science, University of Chinese Academy of Sciences (UCAS), Beijing 100049, China}
\affiliation{Institute for Frontiers in Astronomy and Astrophysics, Beijing Normal University,  Beijing 102206, China}
\affiliation{National Astronomical Observatories, Chinese Academy of Sciences, Beijing 100101, China}

\author[0000-0001-6505-0293]{Keunho J. Kim}
\affiliation{IPAC, California Institute of Technology, Pasadena, CA 91125, USA}

\author[0000-0002-8630-6435]{Anahita Alavi}
\affiliation{IPAC, Mail Code 314-6, California Institute of Technology, 1200 E. California Blvd., Pasadena CA, 91125, USA}

\author[0000-0001-9440-8872]{Norman A. Grogin}
\affiliation{Space Telescope Science Institute, Baltimore, MD 21218, USA}

\author[0000-0003-3759-8707]{Ben Sunnquist}
\affiliation{Space Telescope Science Institute, Baltimore, MD 21218, USA}

\author[0000-0002-0604-654X]{Laura Prichard}
\affiliation{Space Telescope Science Institute, Baltimore, MD 21218, USA}

\author[0000-0003-1268-5230]{Rolf A. Jansen}\affiliation{School of Earth \& Space Exploration, Arizona State University, Tempe, AZ 85287-1404, USA}

\author{the UVCANDELS team}


\shortauthors{Smith, B., et al.} 
\shorttitle{AGN LyC in UVCANDELS}

\begin{abstract}

We present the results of our search for Lyman continuum (LyC) emitting (weak) AGN at redshifts 2.3$\lesssim$z$\lesssim$4.9 from Hubble Space Telescope (HST) Wide Field Camera 3 (WFC3) F275W observations in the UVCANDELS fields. We also include LyC emission from AGN using HST WFC3 F225W, F275W, and F336W found in the ERS and HDUV data. We performed exhaustive queries of the Vizier database to locate AGN with high quality spectroscopic redshifts. In total, we found 51 AGN that met our criteria within the UVCANDELS and ERS footprints. Out of these 51, we find 12 AGN had $\geq$\,4\,$\sigma$ detected LyC flux in the WFC3/UVIS images. Using a wide variety of space-based plus ground-based data, ranging from X-ray to radio wavelengths, we fit the multi-wavelength photometric data of each AGN to a CIGALE SED using AGN models and correlate various SED parameters to the LyC flux. KS-tests of the SED parameter distributions for the LyC-detected and non-detected AGN showed they are likely not distinct samples. However, we find that X-ray luminosity, star-formation onset age, and disk luminosity show strong correlations relative to their emitted LyC flux. We also find strong correlation of the LyC flux to several dust parameters, i.e., polar and toroidal dust emission, 6 $\mu m$ luminosity, and anti-correlation with metallicity and $A_{FUV}$. We simulate the LyC escape fraction (\fesc) using the CIGALE and IGM transmission models for the LyC-detected AGN and find an average \fesc\,$\simeq$\,18\%, weighted by uncertainties. We stack the LyC flux of subsamples of AGN according to the wavelength continuum region in which they are detected and find no significant distinctions in their LyC emission, although our {\it sub-mm detected} F336W sample (3.15\,$<$\,$z$\,$<$\,3.71) shows the brightest stacked LyC flux. These findings indicate that LyC-production and -escape in AGN is more complicated than the simple assumption of thermal emission and a 100\% escape fraction. Further testing of AGN models with larger samples than presented here is needed.

\end{abstract}

\section{Introduction}
The end of the Reionization epoch has been constrained to $z$\,$\sim$\,6 by multiple studies of Gunn-Peterson troughs seen in AGN spectra \citep[e.g.,][]{Becker2001, Djorgovski2001,Fan2006,Goto2011}, though more recent evidence has been presented favoring a longer epoch ending around z\,$\sim$\,5.3--5.5 \citep[e.g.,][]{Qin2021,Bosman2022,Zhu2023}. The leading candidates for the main producers of Hydrogen ionizing radiation ($\lambda$\,$\leq$\,912\,\AA, i.e., Lyman Continuum (LyC)) are thought to be low-mass, star-forming galaxies which existed in abundance at $z$\,$\gtrsim$\,6 \citep[e.g.,][]{Bouwens2015, Parsa2018,SaldanaLopez2023}. However, direct observation of their LyC escape fraction (\fesc) is impossible due to the IGM opacity, and investigations into low-$z$ analogues show the \fesc\ is low \citep[e.g.,][but see, e.g., \citealt{Leitherer2016,Izotov2021,Flury2022} for local LyC emitting analogues]{Rutkowski2017,Marchi2018,Steidel2018,Pahl2021,Griffiths2022}. Furthermore, there is evidence for older stellar populations existing before Reionization completed \citep[e.g.,][]{Hashimoto2018, Laporte2021, Tacchella2023}, indicating that Reionization completion lagged behind significant star formation during galaxy assembly. 

In order to be compatible with the observed rapid Reionization scenario of $\lesssim$\,100\,Myr \citep{Bolan2022}, a second population of low mass, star forming galaxies with high LyC escape must have formed ubiquitously in the Universe from 6\,$\lesssim$\,$z$\,$\lesssim$\,8. These galaxies would require a stellar IMF such that enough supernovae could clear out neutral hydrogen in the ISM via outflows and winds to allow high mass, main sequence stars to emit the sufficient amount of LyC to reionize the IGM with \fesc\,$\simeq$\,10--20\% \citep{Finkelstein2019,Yung2020,Yung2020b,Mutch2023}. These conditions may be able to be met given the rising star-formation histories in some bright $z$\,$>$\,7 galaxies observed with JWST \citep[e.g.,][]{Finkelstein2022,Gimenez2023,Robertson2023,Tacchella2023}. Alternatively, a significant population of low luminosity AGN at $z$\,$>$6 could contribute substantially to the completion of hydrogen Reionization \citep{Giallongo2015,Madau2015,Khaire2016,Grazian2018,Grazian2020,Yung2021,Grazian2022,Onoue2023}. This AGN Reionization paradigm is reinforced by the discovery of $10^8$--$10^9$\,$M_{\odot}$ quasars at $z$\,$>$\,7 when the universe was $\sim$\,750 Myr old \citep{Mortlock2011,Banados2018,Wang2018,Matsuoka2019,Yang2019,Yang2020,Wang2021}, which indicates high efficiency accretion onto super massive black holes (SMBHs) while the IGM was being reionized. Furthermore, recent studies of the $z$\,$\gtrsim$\,5 AGN luminosity function and space density suggest that faint ($M_{1450}$\,$\simeq$\,--23) AGN are $\sim$\,3--5 times more abundant at $z$\,$\sim$\,5 than previous work \citep[e.g.,][]{McGreer2018,Niida2020,Kim2020}, and a slower evolution of the space density from 3\,$\lesssim$\,$z$\,$\lesssim$\,6 is observed \citep[e.g.,][]{Giallongo2019,Grazian2022,Fontanot2023}.

It is most likely the case that stars in massive and low-mass galaxies, as well as AGN, contributed to Reionization. The timeline for which sources dominated the ionizing background, and the characteristics of sources with high \fesc\ remain unclear. In this work, we focus on characterizing the physical properties of AGN with high and low LyC emission using imaging from the HST UVCANDELS program \citep{Wang2023}, which expands on our work from \citet{Smith2018,Smith2020} using WFC3/UVIS ERS \citep{Windhorst2011} and HDUV \citep{Oesch2018} imaging. These constraints can be extrapolated to AGN found during the Reionization epoch, and provide further insight into the role of AGN as reionizers of the IGM, since AGN SED shapes are known to not evolve with cosmic time \citep[e.g.,][]{Shen2019,Yang2021}. We present our analysis as follows: in \S2 we describe the data we use for LyC measurement; in \S3 we detail our sample selection of AGN used for our LyC studies; in \S4 we review our CIGALE SED fitting configuration and the ancillary photometry used for fitting; in \S5 we outline our LyC photometry and \fesc\ methodology; in \S6 we present our results; in \S7 we discuss our results, and in \S8 we summarize our conclusions. We assume \citetalias{Planck2020} cosmology (flat $\Lambda$CDM, $H_0$ = 67.4\,km\,s$^{-1}$\,Mpc$^{-1}$, $\Omega_{m}$\,=\,0.315) and use AB magnitudes \citep{Oke1983} throughout.

\section{Data}
\label{data}
Our main image data used for studying the LyC emission from AGN was collected as part of the Ultraviolet Imaging of the Cosmic Assembly Near-infrared Deep Extragalactic Legacy Survey Fields (UVCANDELS) program \citep{Teplitz2018}. This UV imaging covers the COSMOS, EGS, and completes the GOODS North and South fields in F275W with a 3-orbit depth and coordinated parallels using the F435W filter with the Advanced Camera for Survey (ACS). The F435W exposures have some variation in depth due to roll angle constraints and overlap with other exposures. This data can be found on MAST via \dataset[10.17909/fw24-0b81]{https://doi.org/10.17909/fw24-0b81}. UVCANDELS also includes ground-based Large Binocular Telescope U-band imaging in GOODS North and COSMOS that probes down to $m_{AB}$\,$\sim$\,28\,mag \citep{Ashcraft2018,Ashcraft2023,Otteson2021,Redshaw2022}. We also include UV imaging from the Hubble Deep UV Legacy survey \citep[HDUV;][]{Oesch2018}, the Hubble Ultraviolet Ultra Deep Field \citep[UVUDF;][]{Teplitz2013,Rafelski2015}, and Early Release Science field \citep[ERS;][]{Windhorst2011} in F225W, F275W, and F336W for our AGN LyC analysis when available. The various depths reached by each survey are captured in the photometric uncertainties (see \S\ref{photometry}). Further details on the image quality and drizzling parameters of the UVCANDELS imaging can be found in \citet{Wang2023}. 
\begin{figure*}[ht!]\centerline{
\includegraphics[width=\textwidth]{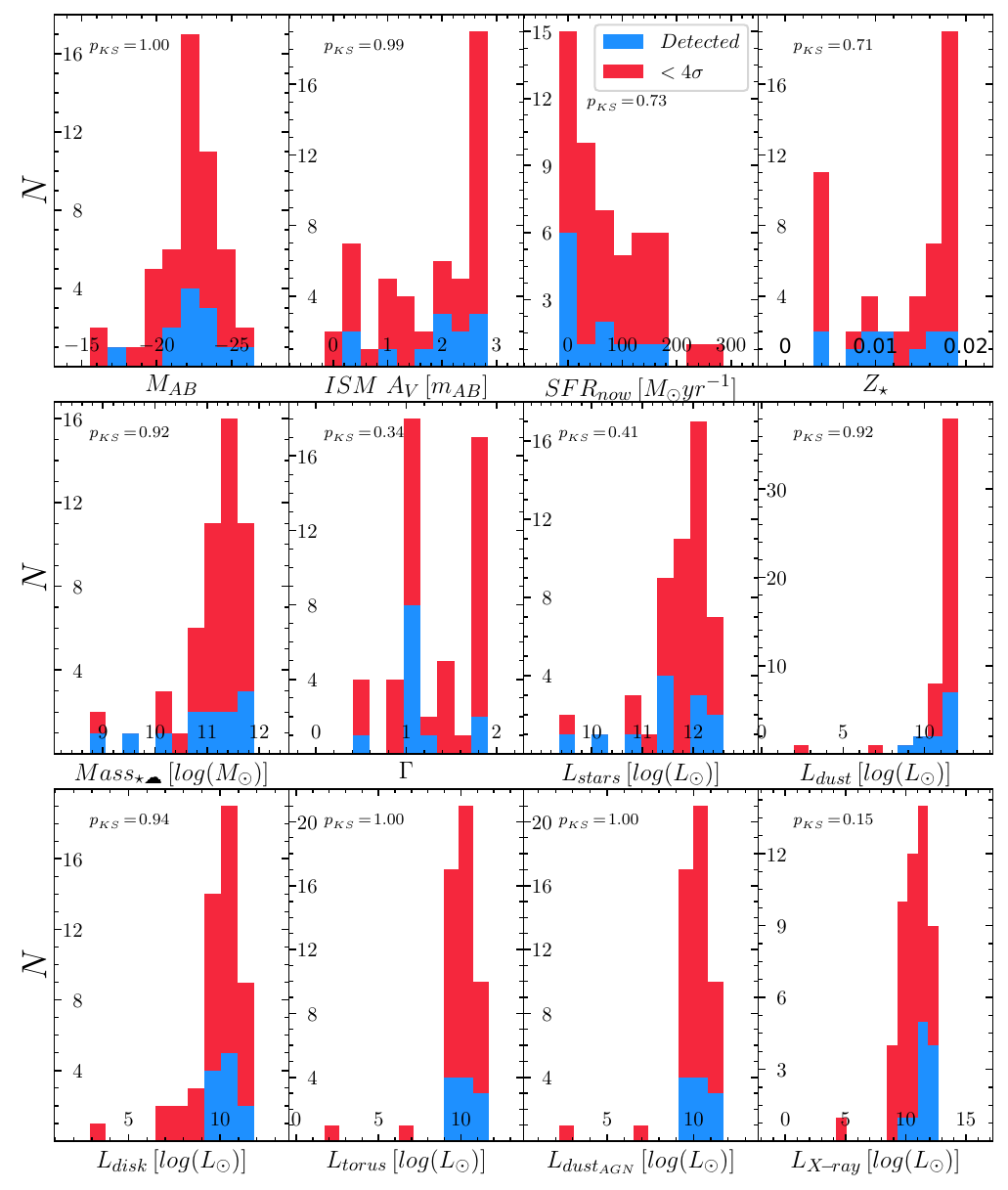}}
\caption{SED parameter histograms of the 51 AGN in our sample. Blue bars indicate AGN with a $\geq$\,4\,$\sigma$ LyC SNR measurement, and red bars represent $<$\,4\,$\sigma$. K-S test of the two histograms cannot reject the null-hypothesis in any histogram. The $M_{AB}$ is calculated from the best-fitting CIGALE SED at $\lambda_{rest}$\,=\,1450\,\AA\ \citep[e.g.,][]{Kulkarni2019}. $A_V$ is taken from the \texttt{Av\_ISM} parameter of the \texttt{attenuation} component, $SFR_{now}$ is from \texttt{sfr} in \texttt{sfh}, $Z$ is from \texttt{metallicity} in \texttt{stellar}, $Mass_{\star\cloud}$ is the sum of \texttt{m\_gas\_old}, \texttt{m\_gas\_young}, \texttt{m\_star\_old}, and \texttt{m\_star\_young} in \texttt{stellar}, $\Gamma$ is from \texttt{gam} in \texttt{xray}, $L_{stars}$ is the sum of \texttt{lum\_old} and \texttt{lum\_young} from \texttt{stellar}, $L_{dust_{abs}}$ is from \texttt{luminosity} in \texttt{dust}, $L_{disk}$ is from \texttt{disk\_luminosity} in \texttt{agn}, $L_{torus}$ is from \texttt{torus\_dust\_luminosity} in \texttt{agn}, $L_{dust_{AGN}}$ is from \texttt{total\_dust\_luminosity} in \texttt{agn}, and $L_{X\!-\!ray}$ is the \texttt{agn\_Lx\_total} parameter in the \texttt{xray} component. \label{hist}}
\end{figure*}

\section{Sample Selection}
\label{sample}
Our spectroscopically verified AGN sample was initially compiled from queries of the VizieR database \citep{Ochsenbein2000} for AGN catalogs and the NASA/IPAC Extragalactic Database (NED)\footnote{The NASA/IPAC Extragalactic Database (NED) is funded by the National Aeronautics and Space Administration and operated by the California Institute of Technology.} for QSOs. LyC from AGN can be safely observed with negligible non-ionizing contamination at redshifts 2.26\,$\leq$\,$z$\,$<$\,2.47, 2.47\,$\leq$\,$z$\,$<$\,3.08, and $z$\,$\geq$\,3.08 in WFC3/UVIS F225W, F275W, and F336W images, respectively \citep{Smith2018}. We therefore selected catalogs and NED objects with spectroscopic redshifts from 2.26\,$\leq$\,$z$\,$<$\,5, and extracted AGN from the selected catalogs \citep{Veron2006,Donley2007,Cardamone2008,Souchay2012,Kirkpatrick2012,Gattano2014,Brightman2014,Souchay2015,Rosen2016,Xue2016,DelMoro2016,Tie2017,Luo2017,Guidetti2017,Liu2018} that were within the footprint of our UV imaging. For each AGN, we further reduced the sample to only include spectra that were deemed the highest quality by the PI of the various spectroscopic surveys \citep{Barger2003,Steidel2003,LeFevre2004,Cowie2004,Szokoly2004,Vanzella2006,Reddy2006,Straughn2008,Barger2008,Popesso2009,Treister2009,Wuyts2009,Balestra2010,Yoshikawa2010,Hodge2012,Kurk2013,Buchner2014,Hsu2014,Brightman2014,Wirth2015,U2015,Albareti2017,Herenz2017,Inami2017,Liu2018,Garilli2021}. We visually inspected the $\sim\!2\farcs0$ region surrounding the AGN in $\chi^2$ images \citep{Szalay1999} we created from all available multi-band \HST\ imaging to reject objects with close ($\lesssim$\,1\farcs5) neighboring objects from our sample. This selection criteria left us with a total of 51 AGN, which we list in table \ref{phottable}.

Our sample exhibits a diverse range of AGN spectral features, seen from the photometry retrieved from NED (see \S\ref{photometry}). In total, we found that 35 AGN showed X-ray emission (Chandra or Swift detected), 50 near-IR emitters ($J$, $H$, or $K$/$K_s$-band), 49 mid-IR emitters (IRAC, WISE 3.4--12$\mu$m, or AKARI S11), 44 far-IR emitters (WISE 22$\mu$m, PACS, IRS, MIPS, or AKARI L18W), 8 sub-mm AGN (SPIRE, SHARC2 350\,$\mu$m, or LABOCA 870\,$\mu$m), 4 microwave AGN (IRAM or AzTEC), and 16 radio AGN (ATCA or VLA). 
\begin{figure*}[ht!]\centerline{
\includegraphics[width=\textwidth]{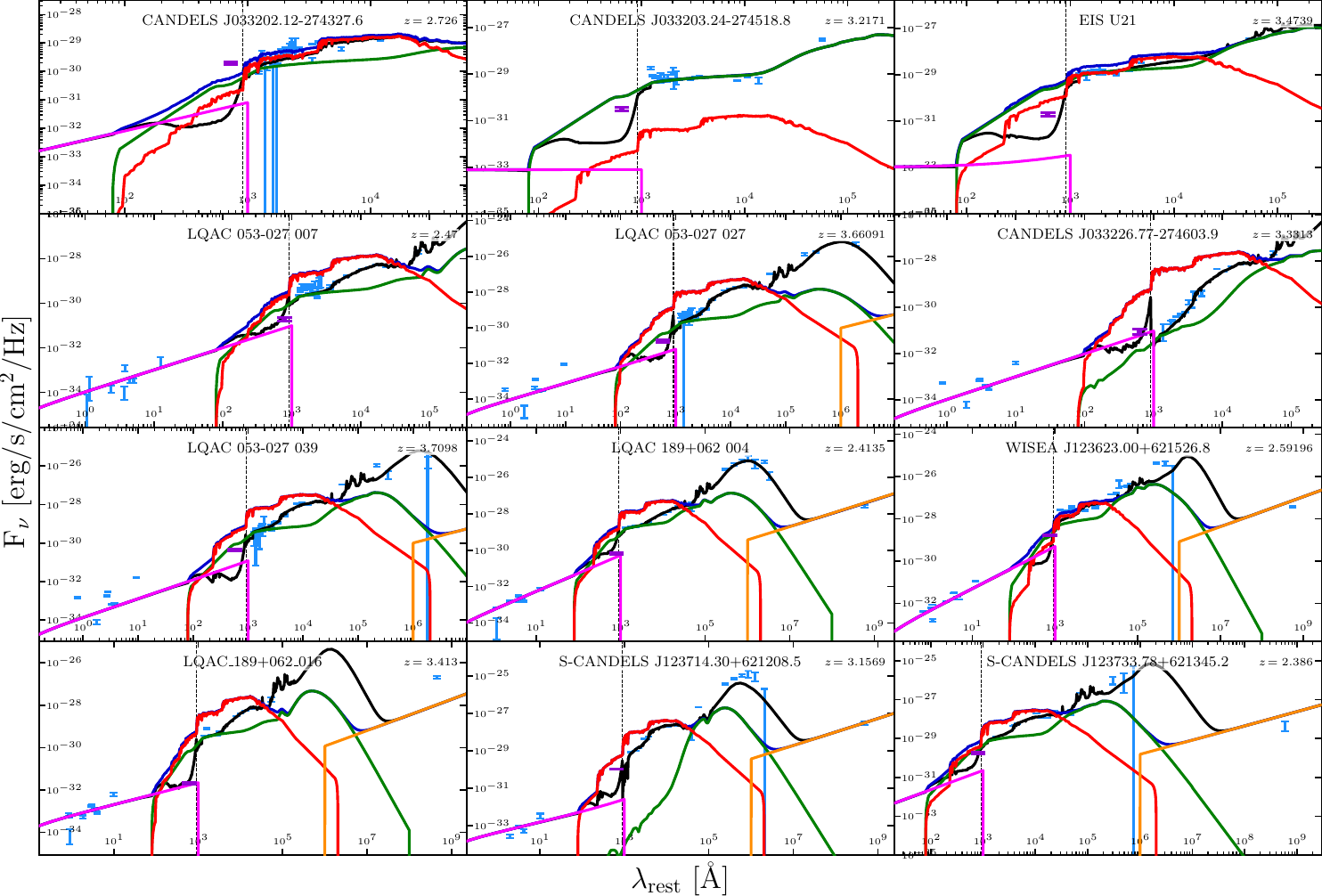}}
\caption{CIGALE SED fitting results for the LyC-detected AGN in table~\ref{detectedTable}. The observed photometry is shown as light-blue error bars, and the LyC detection is indicated as a purple error bar. The fitted SED is shown as a black curve, and the blue curve, used for estimating the escape fraction, includes the same SED components as the black curve, excluding all absorption components. The stellar component is shown as the red curve, the AGN component is shown as the green curve, the X-ray emission is shown as the magenta curve, and the radio emission is shown as the orange curve. }
\end{figure*}

The diversity of our sample is further exemplified in fig.~\ref{hist}. Here we show distributions of several parameters of the best-fitting CIGALE \citep{Boquien2019} SEDs (see \S\ref{sedSec} for a full description of our fitting analysis) with LyC S/N measurements from their respective WFC3/UVIS images of $\geq$\,4\,$\sigma$ and $<$\,4\,$\sigma$ shown as blue and red bars, respectively. We chose 4\,$\sigma$ as the detection threshold since the other 39 AGN below 4\,$\sigma$ showed no obvious LyC flux. The $M_{AB}$ in the first panel is the monochromatic 1450\,\AA\ luminosity calculated from the best-fitting CIGALE SED by integrating the flux between $\lambda_{rest}$\,=\,1449--1551\,\AA\ \citep[e.g.,][]{Kulkarni2019}.

We take the remaining parameters from the physical property estimation variables defined in the CIGALE configuration file. Accordingly, the adopted $A_V$ is taken from the \texttt{Av\_ISM} parameter in the \texttt{attenuation} component of the full SED model, $SFR_{now}$ is defined as the \texttt{sfr} in \texttt{sfh} component, $Z$ is the \texttt{metallicity} parameter in \texttt{stellar} component, $Mass_{\star\cloud}$ is taken as the sum of the values \texttt{m\_gas\_old}, \texttt{m\_gas\_young}, \texttt{m\_star\_old}, and \texttt{m\_star\_young} in \texttt{stellar} component, the X-ray photon index $\Gamma$ is the \texttt{gam} parameter in the \texttt{xray} component, $L_{stars}$ is equal to the sum of the \texttt{lum\_old} and \texttt{lum\_young} parameters from the \texttt{stellar} component, $L_{dust_{abs}}$ is the luminosity absorbed by dust taken from the \texttt{luminosity} parameter in the \texttt{dust} component, $L_{disk}$ is the \texttt{disk\_luminosity} parameter in the \texttt{agn} component, $L_{torus}$ is the \texttt{torus\_dust\_luminosity} parameter in \texttt{agn} component, $L_{dust_{AGN}}$ is the emitted dust luminosity taken from the \texttt{total\_dust\_luminosity} parameter in \texttt{agn} component, and $L_{X\!-\!ray}$ is the \texttt{agn\_Lx\_total} parameter in the \texttt{xray} component. K-S tests of each distribution indicate that the two samples are likely from the same larger sample, as the null hypothesis cannot be rejected. Therefore, high SNR LyC flux emitted from AGN may not be discernable from AGN emitting fainter LyC flux when comparing these parameters. 

Our sample consists of a somewhat fainter population of AGN, all fainter than $M_*$ \citep[$M_{AB}$\,$\simeq$\,--27; see ][]{Kulkarni2019} at the average redshift of $z$\,$\sim$\,3. Their SED fits exhibit dusty, weak-AGN characteristics, with the stellar component being the dominant source of their intrinsic luminosity for $\sim$\,35\% of the sample, while dust dominates the luminosity for $\sim$\,45\%. Only $\sim$\,4\% of the sample has the AGN disk component as their brightest feature. \citet{Grazian2018} and \citet{Romano2019} performed a similar study to our present work on AGN with --25.1\,$\leq$\,$M_{UV}$\,$\leq$\,--23.3 and --29.0,$\leq$\,$M_{UV}$\,$\leq$\,--26.0, respectively. Most of their AGN are brighter than ours, which likely explains the ubiquity of detections amongst their samples. Thus, this work serves as a complement to the study of more luminous quasar LyC detections, which do not include LyC non-detected AGN. Their escape fractions are also much higher than the present sample, even for \emph{our} LyC-detected AGN (see \S\ref{detectedTable}), with the lowest \fesc\,=\,44\% from the \citet{Grazian2018} sample. Compared to the \citet{Wang2023} galaxy sample, 19/51 AGN host-galaxies lie in the same range of --22$\,$\,$M_{UV}$\,$<$--18, with four of these being LyC-detected. Here, $M_{UV}$ is the monochromatic 1500\,\AA\ absolute magnitude calculated from the best-fit CIGALE SEDs, excluding the AGN components. Eight AGN host-galaxies are brighter than $-22$\,Mag, two of which are LyC-detected, and 24 are fainter than $-18$\,Mag, six of which are LyC-detected. This likely indicates that the AGN itself is emitting most of the detected LyC rather than the host-galaxy's massive stars. Compared to the \citet{Steidel2018} sample, the monochromatic 1700\,\AA\ luminosity of the SEDs from our sample without the AGN component mostly fall within the same range of $L_{UV}/L^*_{UV}$\,$<$3 ($L^*_{UV}$\,=\,--21 for $z$\,$\simeq$\,3; \citealt{Reddy2009}). However, we find that 7/51 AGN host-galaxy stellar components exceed this ratio, by an order of magnitude in three cases. Two of these were detected in LyC, namely S-CANDELS\,J123714.30+621208.5 and WISEA\,J123623.00+621526.8, with a ratio of $\sim$\,9 and $\sim$\,20 respectively. WISEA\,J033209.44-274807.3 had an exceptional $M_{AB_{1700\mathrm{\AA}}}$\,$\simeq$\,--25.13 for its non-AGN SED components, though it was one of the four bright non-detections. 

The X-ray luminosity outshines all other SED components for only $\sim$\,16\% of the sample, indicative of mostly obscured AGN, and all X-ray detected AGN display a soft photon index ($\Gamma$\,$<$\,2). The population also shows significant star formation, with 50\% of the sources showing a SFR between 10--100 $M_{\odot}\,/yr$, and $\sim$\,37\% at SFR\,$>$\,100\,$M_{\odot}\,/yr$, with only the remaining 10\% at SFR\,$<$\,10\,$M_{\odot}\,/yr$. Metallicities are more diverse, with $\sim$\,32\%\ showing $Z$\,$<$\,0.01, and $\sim$\,28\%\ having $Z$\,$\geq$\,0.02, and the rest of the sample having 0.01\,$\leq$\,$Z$\,$\leq$\,0.02. 

\section{SED Fitting\label{sedSec}}
\subsection{Photometry}
\label{photometry}
The photometry used for SED fitting was compiled from region-based queries of the NED database. We compiled all available photometry within 1\farcs5 of the centroid of the corresponding $\chi^2$ image mentioned in \S\ref{sample}. This position offset tolerance was chosen so that our query would capture photometry with larger PSFs, e.g., from X-ray and ground-based data. We homogenized the retrieved photometry to units of mJy based on the instrument and band for all available photometric data. We then added the photometric data point with the highest S/N for each band to our photometric catalog if there was more than one measurement for a particular band. We also filtered out photometric bands from our query that did not have available transmission curves to be used for SED fitting. \vspace{-2\baselineskip}

\tabletypesize{\footnotesize}
\begin{deluxetable*}{lllclr}
\tablecaption{Individual LyC Detections\label{detectedTable}}
\tablehead{
\colhead{\small ID} & \colhead{\small $z$} & \colhead{\small m$_{\!\lyc}$} & \colhead{\small SNR$_{\!\lyc}$} & \colhead{\small \fesc} & \colhead{\small m$_{\uvc}$} \\[-5pt]
\colhead{} & \colhead{} & \colhead{[$m_{AB}$]} & \colhead{} & \colhead{[\%]} & \colhead{[$m_{AB}$]}\\[-5pt]
\colhead{\small (1)} & \colhead{\small (2)} & \colhead{\small (3)} & \colhead{\small (4)} & \colhead{\small (5)} & \colhead{\small (6)}
}
\startdata
CANDELS J033202.12-274327.6 & 2.726 & 25.68$\pm$0.11 & 10.04 & 227$\pm$83.5 & 25.65$\pm$0.04 \\[-1pt]
CANDELS J033203.24-274518.8 & 3.2171 & 27.64$\pm$0.17 & 6.27 & 17$\pm$6.7 & 24.82$\pm$0.01 \\[-1pt]
EIS U21 & 3.4739 & 28.14$\pm$0.20 & 5.41 & 5$\pm$1.6 & 24.14$\pm$0.00 \\[-1pt]
LQAC 053-027 007 & 2.47 & 28.16$\pm$0.21 & 5.21 & 7$\pm$2.4 & 25.75$\pm$0.01 \\[-1pt]
LQAC 053-027 027 & 3.6609 & 28.24$\pm$0.17 & 6.25 & 1$^{+8.1}_{-1}$ & 25.69$\pm$0.02 \\[-1pt]
CANDELS J033226.77-274603.9 & 3.3313 & 29.04$\pm$0.27 & 3.99 & 4$^{+29.2}_{-4}$ & 27.23$\pm$0.15 \\[-1pt]
LQAC 053-027 039 & 3.7098 & 27.28$\pm$0.11 & 9.71 & 7$^{+44.2}_{-7}$ & 24.85$\pm$0.01 \\[-1pt]
LQAC 189+062 004 & 2.4135 & 26.86$\pm$0.07 & 16.06 & 3$^{+13.4}_{-3}$ & 23.66$\pm$0.00 \\[-1pt]
WISEA J123623.00+621526.8 & 2.592 & 23.35$\pm$0.00 & 585.77 & 65$\pm$2.6 & 20.77$\pm$0.00 \\[-1pt]
LQAC 189+062 016 & 3.413 & 30.60$\pm$0.11 & 9.88 & 2$\pm$6.9 & 26.42$\pm$0.01 \\[-1pt]
S-CANDELS J123714.30+621208.5 & 3.1569 & 26.31$\pm$0.00 & 375.68 & 101$\pm$9.0 & 25.97$\pm$0.01 \\[-1pt]
S-CANDELS J123733.78+621345.2 & 2.386 & 25.75$\pm$0.14 & 7.72 & 12$^{+124.7}_{-12}$ & 23.66$\pm$0.01 \\[-1pt]
\enddata
\tablecomments{\textbf{Table columns:} (1) NED object ID; (2) AGN spectroscopic redshift; (3) LyC AB magnitude measured in the corresponding WFC3/UVIS filter; (4) S/N of measured LyC flux; (5) The \fesc\ of the AGN and host-galaxy and its uncertainties. Values over 100\%\ are included to show the limitations of the SED models.; (6) The non-ionizing UV AB magnitude measured from the ACS/WFC F606W filter.}
\end{deluxetable*}

The resulting photometric catalog included data from the 2.2-m MPG/ESO WFI (B, V R, and I), 2MASS J \& K$_s$, AKARI S11 \& L18W, ATCA CABB and 1.4 GHz bands, AzTEC 1.16\,mm, Blanco/DECam ($g$, $r$, $i$, $z$, and Y), Blanco/MOSAIC-II H-$\alpha$, CFHT/WIRCam J \& K, Chandra, COMBO-17, Gemini/QUIRC H+K, Hale/WIRC J \& K$_s$, Herschel (PACS and SPIRE), HST (ACS/WFC F435W, F606W, F775W, F814W, and F850LP, WFC3/IR F098M, F105W, F125W, F140W, and F160W, ACS/SBC F150LP, WFPC2 F300W, and NICMOS F110W and F160W), IRAM 1.2\,mm and 2.2\,mm, LABOCA 870\,$\mu$m, KPNO/Mayall (G and R from \citet{Steidel2003}), SHARC2 350\,$\mu$m, Spitzer (IRAC, IRS, and MIPS), NTT/SOFI (J, H, and K$_s$), SDSS $griz$, SCUBA-2 850\,$\mu$m, Subaru/Suprime-Cam (B, V, R, I, and $z$), Swift, VISTA/VIRCAM (Y, J, H, and K$_s$), VLA (L and C-band), VLT/ISAAC (J, H, and K$_s$-band), VLT/HAWK-I K$_S$, WISE, and XMM-OM (B and V). This was compiled from a large sample of photometric catalogs: \citet{Brandt2001,Wolf2001,Alexander2003,Steidel2003,Abazajian2004,Chen2004,Cowie2004,Jahnke2004,Koekemoer2004,LeFevre2004,Smail2004,Wirth2004,Wolf2004,Akiyama2005,Borys2005,Bouwens2005,Pope2005,Rigby2005,Thompson2005,Afonso2006,AlonsoHerrero2006,Beckwith2006,Biggs2006,Coe2006,Kovacs2006,Laird2006,Reddy2006,Steffen2006,Weiner2006,Kelly2007,Rodighiero2007,AdelmanMcCarthy2008,Aird2008,Cardamone2008,Finkelstein2008,Kellermann2008,Mainieri2008,Miller2008,Wiklind2008,Wuyts2008,Chen2009,Daddi2009,Mancini2009,Negrello2009,Rigopoulou2009,Treister2009,Young2009,Carilli2010,Lutz2010,Morrison2010,Pearson2010,Cameron2011,Georgantopoulos2011,Magnelli2011,Penner2011,Puccetti2011,Reis2011,Shim2011,Teplitz2011,Vagnetti2011,Cutri2013,Bizzocchi2014,Bouwens2014,Cutri2014,Huang2014,Skelton2014,Straatman2014,Tan2014,Antonucci2015,Ashby2015,Franzen2015,Falocco2015,Huynh2015,JimenezTeja2015,Cappelluti2016,Harikane2016,Wang2016,Xue2016,Cowie2017,Guidetti2017,Tie2017,Liu2018,Vito2018}. All of these catalogs were used for SED fitting for at least one AGN, but not all AGN had photometry taken from every catalog. The brightest band and its flux in each part of the AGN emission spectrum is listed in the Appendix table~\ref{phottable}.

\subsection{CIGALE Configuration}
We used the CIGALE \citep{Boquien2019} SED fitting code to model the AGN emission from X-ray to radio wavelengths. These models enable us to calculate the intrinsically produced LyC of the AGN and host-galaxy. We used the $\texttt{sfhdelayedbq}$ model and $\texttt{bc03}$ templates to model the star-formation history and stellar populations, $\texttt{nebular}$ templates to simulate emission lines from the ISM, $\texttt{dustatt\_modified\_CF00}$ model to simulate dust attenuation, $\texttt{skirtor2016}$ templates to fit the AGN emission, $\texttt{dale2014}$ templates to simulate the thermal dust emission, $\texttt{xray}$ model to simulate the X-ray emission from stars and the AGN, and the $\texttt{radio}$ module model to simulate the galaxy synchrotron and AGN emission during fitting. 

These components are a combination of mathematical models with free parameters and templates with pre-defined choices of parameters, i.e. the $\texttt{bc03}$, $\texttt{nebular}$, $\texttt{dale2014}$, and $\texttt{skirtor2016}$ templates. All but the $\texttt{dale2014}$ templates, however, are in-turn based on physically motivated, empirical mathematical models \citep{Ferland1998,Ferland2013,Bruzual2003,Stalevski2012,Stalevski2016}. These templates, of course, limit the parameter space by the resolution and range of the parameters in the templates provided by CIGALE. Incorporating additional models not included in the main distribution of CIGALE could improve this. Another limitation we observed comes from how the $\texttt{nebular}$ component scales the LyC to provide energy to emission lines. When the observed LyC flux exceeds the modeled LyC flux (which includes a wavelength-dependent IGM attenuation factor at redshift $z$), the $\texttt{nebular}$ LyC scaling may produce erroneous results. This case can occur for high \fesc\ galaxies and AGN, since higher \fesc\ values have likely been observed through lines-of-sight with higher than average IGM transmission. When including photometry below the rest-frame Lyman-limit, the modeled LyC gets scaled to meet this observed LyC flux, creating an unphysical ``jump'' or discontinuity at 912\,\AA. To circumvent this, we did not fit any SEDs to photometry measured at rest-frame wavelengths shorter than 1216\,\AA.

We also modified the code to extrapolate the power law fit of the X-ray model to 1000\,\AA\ specifically for our AGN since the AGN continuum power-law shows a break at around this wavelength \citep{Telfer2002}, and also to fill the discontinuity in the model between 220\,$\lesssim$\,\AA\,$\lesssim$\,400 that exists without this modification. This will also account for the hard UV tail of the AGN warm corona emission \citep[e.g.,][]{Lusso2015,Petrucci2020}.
\begin{figure*}[ht!]\centerline{
\includegraphics[width=\textwidth]{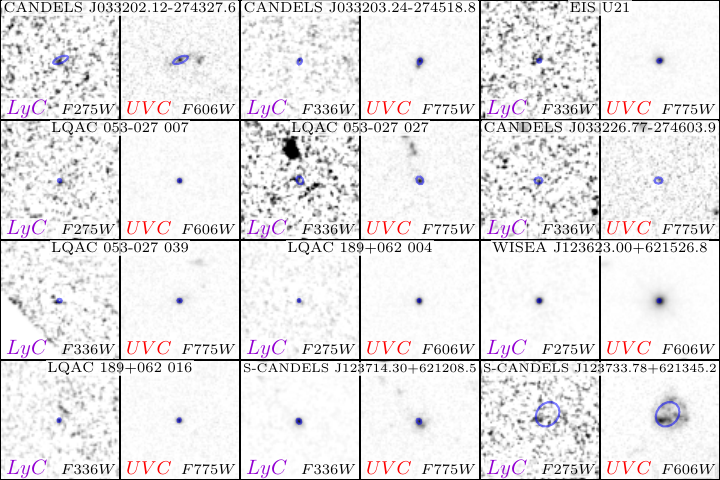}}
\caption{4\farcs53$\times$4\farcs53 cutouts of our LyC detected AGN showing their LyC image (LyC, left) and $\sim$\,1500\AA\ rest-frame UV continuum (UVC, right). The NED object name and corresponding filters are also shown. The $\chi^2$ image detected apertures are shown as blue ellipses. \label{detections}}
\end{figure*}

\section{LyC Photometry and Escape Fractions}

We performed all LyC photometry as we have in our previous LyC work from \citetalias{Smith2018} and \citetalias{Smith2020}. In summary, we treat each pixel as a random variable with normally distributed flux, where the mean is set to the pixel value from the drizzled science image, and the dispersion is set to the RMS value from the corresponding pixel in the weightmap, plus the local sky dispersion measured in the 151\,$\times$\,151 pixels (4\farcs53\,$\times$\,4\farcs54) surrounding the target, normalized to each individual pixel. Neighboring object pixels were also masked in this process. We then generated 10,000 randomly drawn 151\,$\times$\,151\,pixel images using these statistics, and measured the flux within the Kron-like elliptical aperture detected from the $\chi ^2$ image for each realized image \citep[created using all available \HST\ data; see][]{Szalay1999} using SExtractor \citep{Bertin1996}. The resulting flux distribution from these measurements was used to determine the LyC flux and its uncertainties given in table \ref{detectedTable}. When performing photometric measurements on our LyC stacks (see \S\ref{lycstacking} for details on our stacking procedure), we treated each pixel in each sub-image of a stack as a random variable which creates a 151\,$\times$\,151\,$\times$\,10,000 data cube for that sub-image, then stacked each data cube using a weighted sum. We then normalized the stack of data cubes by a stack of all weightmap cutouts. The flux distribution was then generated in the same way as the individually detected AGN, taking slices from the data cube and measuring the central isophotal flux, using the $\chi^2$ image for detection. The resulting stacked photometry and its uncertainties is presented in more detail \S\ref{lycstacking} and table~\ref{stackPhotometryTable}.

To calculate the LyC \fesc, we used the same method as in \citetalias{Smith2018} and \citetalias{Smith2020}. In our method, we simply take the ratio of the observed LyC flux to an intrinsic LyC flux modelled using the unattenuated SED, the WFC3/UVIS filter curve corresponding to the observed LyC flux filter, and IGM attenuation curves modeled at the redshift of the target AGN. In short, we modeled the intrinsic LyC flux from the best-fitting CIGALE SED using only the following components: \texttt{stellar.old}, \texttt{stellar.young}, \texttt{nebular.lines\_young}, \texttt{nebular.lines\_old}, \texttt{nebular.continuum\_young}, \texttt{nebular.continuum\_old}, \texttt{agn.SKIRTOR2016\_}

\noindent\texttt{disk}, \texttt{agn.SKIRTOR2016\_torus}, \texttt{agn.SKIRTOR-}

\noindent\texttt{2016\_polar\_dust}, \texttt{xray.galaxy}, \texttt{xray.agn}, \texttt{radio.sf\_nonthermal}, and \texttt{radio.agn}. We summed these components to acquire our intrinsic SED, then we calculated the inner product of this SED with the IGM attenuation curves and filter transmission curves to obtain our final intrinsic LyC flux model. We performed this inner product 10,000 times for the variety of sight-lines through the IGM at the redshift of the AGN using the model from \citet{Inoue2014}. We used the flux distribution described in the previous paragraph as the observed LyC flux, took the ratio of these two distributions, and used this ratio to calculate the \fesc\ distribution and its statistics which are presented in table~\ref{detectedTable}. It is worth noting that the modeled LyC from the SED, as well as the apertures shown in Fig~\ref{detections} are intended to capture both the LyC emission from the central AGN as well as any stellar LyC leaking from the host galaxy.

We find a wide range of \fesc\ values from 1--100\%\ or greater, which represent the mode of the \fesc\ distribution. Corresponding 1$\sigma$ uncertainties, along with the \fesc\ values are shown in Table~\ref{detectedTable}. An inverse variance weighted average value including all values in Table~\ref{detectedTable} (with a ceiling of 100\%) equates to approximately \fesc\,$\simeq$\,18\%. \fesc\ distributions with modes above 100\%\ result from higher observed LyC compared to the modeled, intrinsic LyC obtained from the best-fit CIGALE SED. Although these \fesc\ values are unphysical, and at a $\sim$2.7$\sigma$ significance, we show the \fesc\ in these cases to highlight the need for AGN models that can accommodate high LyC emission. 

\subsection{LyC Stacking}
\label{lycstacking}
Stacking is a technique generally used to increase the SNR of signals from sources of interest, or to find the average signal of a sample. In this work, we stack LyC flux from galaxies hosting AGN to compare the total LyC signal brightness of various subsets within our sample based on observed spectral features. To distinguish these subsets, we used the archival photometry retrieved from the various catalogs hosted on Vizier and categorized the detected photometric bands into regions of the electromagnetic spectrum, namely X-ray, optical, IR, MIR, FIR, sub-mm, microwave, and radio. We did not include a UV-detected sample since U-band may overlap with the rest-frame LyC for some higher-$z$ AGN. Each of these subsamples is defined by having a detection in their respective region of the EM spectrum. The bandpass filters used for determining the inclusion in each subsample can be found in \S\ref{photometry}. 
\begin{figure*}[ht!]\centerline{
\includegraphics[width=\textwidth]{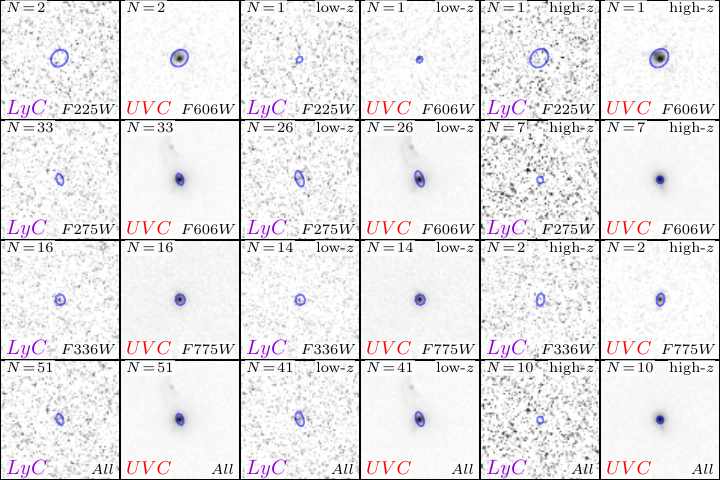}}
\caption{\label{stacks}LyC and UVC stacks of all AGN in the total sample. Blue ellipses are SExtractor detection apertures from the UVC stack. The number of AGN included in each stack is indicated on the top-left corner of each individual panel. Left two columns show the full sample of AGN where LyC can be measured in either the filter indicated on the bottom right of the panel, or the total AGN (``All'') sample in the bottom row. The middle and right pair of columns represent half of the AGN in their row, split into low and high redshift samples using the average redshift of the sub-sample as a separator. Top row: AGN with LyC measurable in F225W; Top-middle row: same but for F275W; Top-bottom row: same but for F336W; Bottom row: Same, but for all AGN in the total sample.}
\end{figure*}

We show the resulting stacks of galaxies in each band in fig.~\ref{stacks}, and the photometry performed for each stack is presented in table~\ref{stackPhotometryTable}. We find the stack with the brightest LyC emission is the sub-mm detected AGN in the F336W stack, with redshifts that range from 3.15\,$<$\,$z$\,$<$\,3.71. There are only three objects in this stack, so it is likely there is a detection bias since detection in sub-mm often implies overall brighter luminosity across the SED. The LyC stack of all AGN detected in a particular region of the electromagnetic spectrum was the IR- and MIR-detected AGN LyC stacks, which includes 50 and 49 out of the 51 AGN, respectively. The AGN missing from these stacks simply did not have any IR/MIR band observations. A zero-point normalized stack (the ``All LyC'' row in the All AGN set of stacks in table~\ref{stackPhotometryTable}) was brighter than these two stacks, and includes all 51 AGN. We also stack the LyC non-detections in each of our three redshift bins, as well as a stack of all non-detections. None of these stacks resulted in detected LyC.

These results demonstrate that stacks based on detected regions in the AGN electromagnetic continuum do not correlate well with their LyC emission. There are many factors that can affect the LyC flux that is detected, e.g., sample size statistics, IGM line of sight attenuation, AGN inclination, observed filter and its proximity to the Lyman-limit, or other intrinsic physical factors of the host-galaxy or AGN. A correlation analysis on an individual basis of each AGN's LyC emission may reveal more clues for which regions of the continuum could indicate LyC leakage. 

We also created stacks based on the filter-cutoff's proximity to the AGN's observed-frame Lyman-limit. We stacked the LyC cutouts in two redshift bins (low-$z$ and high-$z$) separated by their median redshift for four samples in the ``All AGN'' set of stacks, i.e., the F225W, F275W, F336W, and ``All'' samples. In all cases, we see that the low-$z$ substacks, which would have their Lyman-limit closer to the filter cut-off, show brighter LyC emission than the high-$z$ substacks. This result is not surprising since a given filter would observe shorter wavelength flux for higher redshift galaxies, and the central AGN and stellar sources produce less LyC flux further from the Lyman-limit. From this, we conclude that LyC observations should attempt to select targets with Lyman-limits as close as possible to their filter-cutoffs, while minimizing red-leak in the filter beyond the Lyman-limit below some fiducially low percentage.

Of the stacks based on redshift ranges which correspond to observed WFC3/UVIS F225W, F275W, and F336W filters, we find that the stack of all galaxies in the F336W sample (3.08\,$<$\,$z$\,$<$\,4.88 \zmean\,$\sim$\,3.5924) containing 16 AGN was the brightest substack with \mAB\,=\,27.66$\pm$0.56. This result is more interesting, since the higher redshift AGN should experience more attenuation from the IGM. Furthermore, AGN activity peaks at $z$\,$\sim$\,2--3 in space-density and $M_*$ \citep{Kulkarni2019}, and AGN in this stack are expected to be intrinsically fainter. Since AGN do not exhibit evolution in their SED through redshift \citep[e.g.,][]{Barth2003,Fan2004,Iwamuro2004,Jiang2007}, the LyC emission from AGN may not be redshift dependent either. Thus, it is more likely the case that LyC emission from AGN is solely dependent on physical parameters of the AGN and host-galaxy, or inhibited to some degree by the surrounding IGM. 

\begin{figure*}[ht!]\centerline{
\includegraphics[width=\textwidth]{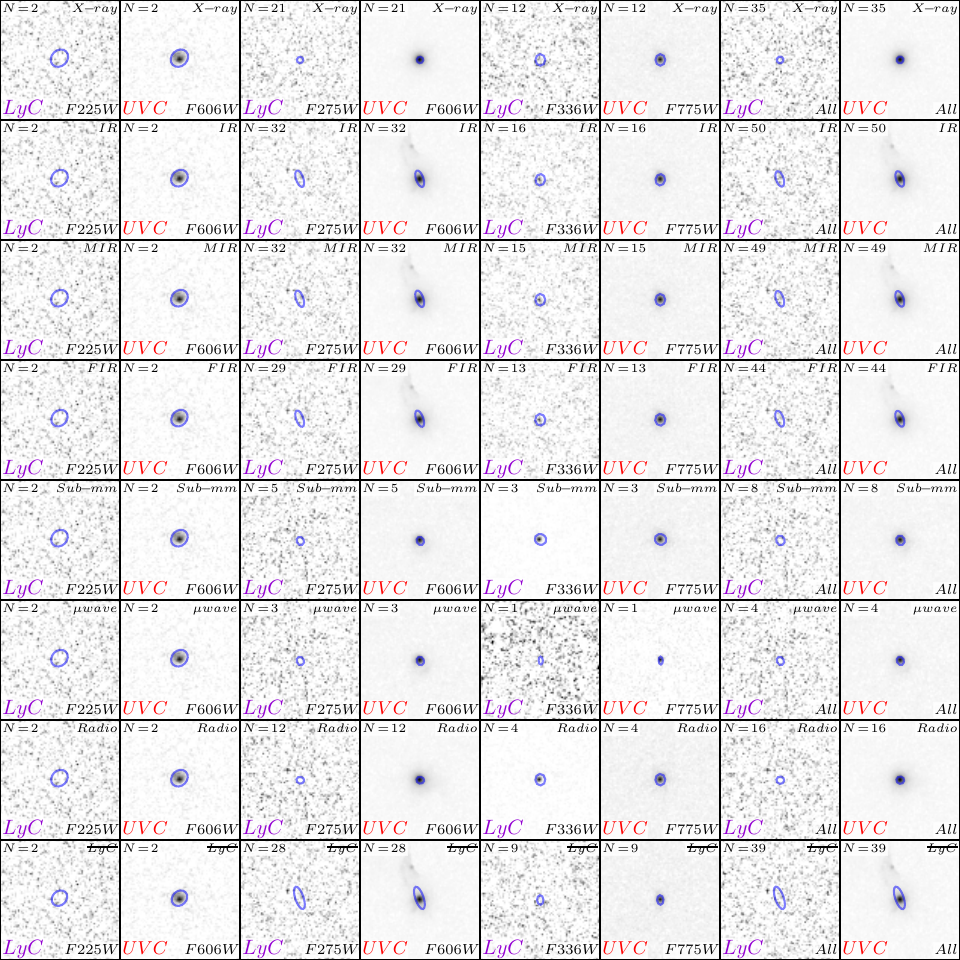}}
\caption{LyC and UVC stacks of AGN sub-samples listed in table~\ref{stackPhotometryTable}. Top row: Sub-sample of AGN that were detected by Chandra and/or Swift; 2nd row: AGN detected in an IR band (J, H, or K); 3rd row: AGN detected in a Mid-IR band (IRAC, WISE 3.4--12$\mu$m, or AKARI S11); 4th row: AGN detected in a Far-IR band (WISE 22$\mu$m, PACS, IRS, MIPS, or AKARI L18W); 5th row: AGN detected in a sub-mm band (SPIRE, SHARC2 350\,$\mu$m, or LABOCA 870\,$\mu$m); 6th row: AGN detected in the microwave region (IRAM or AzTEC); 7th row: AGN detected in the radio (ATCA or VLA). Bottom row: LyC non-detected AGN, indicated by \sout{$LyC$}. The IR-detected AGN show the brightest flux among the "All" sub-samples, with 39 included in the stack. The brightest sub-sample of all panels is the F336W IR-detected sub-sample. }
\end{figure*}

\begin{deluxetable*}{lcccrccclr}
\tablecaption{Lyman Continuum Stack Photometry\label{stackPhotometryTable}}
\tablehead{
\colhead{Filter} & \colhead{$z$-range} & \colhead{\zmean} & \colhead{$N_{obj}$} & \colhead{m$_{\lyc}$} & \colhead{ABerr$_{\!\lyc}$} & \colhead{S/N$_{\lyc}$} & \colhead{A$_{\uvc}$} & \colhead{m$_{\uvc}$} & \colhead{ABerr$_{\uvc}$}\\[-5pt]
\colhead{(1)} & \colhead{(2)} & \colhead{(3)} & \colhead{(4)} & \colhead{(5)} & \colhead{(6)} & \colhead{(7)} & \colhead{(8)} & \colhead{(9)} & \colhead{(10)}
}
\startdata
\multicolumn{10}{l}{\sc All AGN:}\\
F225W & 2.2807-2.2979 & 2.2893 & 2 & $>$27.01 & \nodata & $(1.0)^{\ddagger}$ & 1.28 & 25.626 & 0.024 \\ [-4pt]
F225W (low-$z$) & 2.2807-2.2807 & 2.2807 & 1 & $>$29.26 & \nodata & $(1.0)^{\ddagger}$ & 0.19 & 27.994 & 0.112 \\ [-4pt]
F225W (high-$z$) & 2.2979-2.2979 & 2.2979 & 1 & $>$26.92 & \nodata & $(1.0)^{\ddagger}$ & 1.27 & 25.017 & 0.019 \\ [-4pt]
F275W & 2.3860-3.0750 & 2.6117 & 33 & 28.83 & 0.62 & 1.75 & 0.57 & 23.314 & 0.000 \\ [-4pt]
F275W (low-$z$) & 2.3860-2.7280 & 2.5324 & 26 & 28.19 & 0.56 & 1.95 & 0.98 & 23.654 & 0.001 \\ [-4pt]
F275W (high-$z$) & 2.7370-3.0750 & 2.9062 & 7 & $>$29.29 & \nodata & $(1.0)^{\ddagger}$ & 0.20 & 22.681 & 0.000 \\ [-4pt]
F336W & 3.0804-4.8759 & 3.5924 & 16 & 27.66 & 0.56 & 1.95 & 0.50 & 24.941 & 0.004 \\ [-4pt]
F336W (low-$z$) & 3.0804-3.9050 & 3.4445 & 14 & 29.07 & 0.31 & 3.53 & 0.46 & 24.876 & 0.004 \\ [-4pt]
F336W (high-$z$) & 4.3792-4.8759 & 4.6275 & 2 & $>$28.95 & \nodata & $(1.0)^{\ddagger}$ & 0.57 & 25.878 & 0.031 \\ [-4pt]
All LyC & 2.2807-4.8759 & 2.9067 & 51 & 28.46 & 0.45 & 2.39 & 0.54 & 23.874 & 0.001 \\ [-4pt]
All LyC (low-$z$) & 2.2807-3.9050 & 2.8377 & 41 & 30.51 & 0.55 & 1.96 & 0.84 & 24.252 & 0.001 \\ [-4pt]
All LyC (high-$z$) & 2.2979-4.8759 & 3.1896 & 10 & $>$29.09 & \nodata & $(1.0)^{\ddagger}$ & 0.21 & 23.053 & 0.000 \\ 
\midrule\\[-13pt]
\multicolumn{10}{l}{\sc LyC non-detections:}\\
F225W & 2.2807-2.2979 & 2.2893 & 2 & $>$27.01 & \nodata & $(1.0)^{\ddagger}$ & 1.28 & 25.626 & 0.024 \\ [-4pt]
F275W & 2.4030-3.0750 & 2.6285 & 28 & $>$28.41 & \nodata & $(1.0)^{\ddagger}$ & 0.87 & 23.730 & 0.001 \\ [-4pt]
F336W & 3.0804-4.8759 & 3.7239 & 9 & 27.93 & 0.47 & 2.31 & 0.28 & 25.437 & 0.007 \\ [-4pt]
All LyC & 2.2807-4.8759 & 2.8639 & 39 & 28.83 & 0.91 & 1.20 & 0.89 & 24.121 & 0.001 \\ 
\midrule\\[-13pt]
\multicolumn{10}{l}{\sc Xray detected:}\\
F225W & 2.2807-2.2979 & 2.2893 & 2 & $>$26.95 & \nodata & $(1.0)^{\ddagger}$ & 1.34 & 25.623 & 0.024 \\ [-4pt]
F275W & 2.4135-3.0750 & 2.6613 & 21 & $>$29.16 & \nodata & $(1.0)^{\ddagger}$ & 0.18 & 22.943 & 0.000 \\ [-4pt]
F336W & 3.0804-4.8759 & 3.5972 & 12 & 27.89 & 0.77 & 1.42 & 0.50 & 25.033 & 0.005 \\ [-4pt]
All LyC & 2.2807-4.8759 & 2.9609 & 35 & 29.47 & 0.78 & 1.39 & 0.21 & 23.613 & 0.000 \\ 
\midrule\\[-13pt]
\multicolumn{10}{l}{\sc IR detected:}\\
F225W & 2.2807-2.2979 & 2.2893 & 2 & $>$27.15 & \nodata & $(1.0)^{\ddagger}$ & 1.24 & 25.662 & 0.024 \\ [-4pt]
F275W & 2.3860-3.0750 & 2.6105 & 32 & 28.93 & 0.72 & 1.51 & 0.56 & 23.304 & 0.000 \\ [-4pt]
F336W & 3.0804-4.8759 & 3.5924 & 16 & 27.51 & 0.36 & 3.01 & 0.49 & 24.949 & 0.004 \\ [-4pt]
All LyC & 2.2807-4.8759 & 2.9119 & 50 & 28.66 & 0.54 & 2.03 & 0.55 & 23.869 & 0.001 \\ 
\midrule\\[-13pt]
\multicolumn{10}{l}{\sc MIR detected:}\\
F225W & 2.2807-2.2979 & 2.2893 & 2 & $>$27.14 & \nodata & $(1.0)^{\ddagger}$ & 1.24 & 25.666 & 0.024 \\ [-4pt]
F275W & 2.3860-3.0750 & 2.6104 & 32 & 29.09 & 0.81 & 1.35 & 0.57 & 23.282 & 0.000 \\ [-4pt]
F336W & 3.0804-4.8759 & 3.5842 & 15 & 27.73 & 0.52 & 2.10 & 0.52 & 24.878 & 0.004 \\ [-4pt]
All LyC & 2.2807-4.8759 & 2.8954 & 49 & 28.66 & 0.62 & 1.75 & 0.55 & 23.835 & 0.001 \\ 
\midrule\\[-13pt]
\multicolumn{10}{l}{\sc FIR detected:}\\
F225W & 2.2807-2.2979 & 2.2893 & 2 & $>$27.05 & \nodata & $(1.0)^{\ddagger}$ & 1.20 & 25.716 & 0.025 \\ [-4pt]
F275W & 2.3860-3.0750 & 2.6199 & 29 & $>$28.84 & \nodata & $(1.0)^{\ddagger}$ & 0.55 & 23.181 & 0.000 \\ [-4pt]
F336W & 3.0804-4.3792 & 3.4926 & 13 & 27.60 & 0.43 & 2.55 & 0.53 & 24.795 & 0.004 \\ [-4pt]
All LyC & 2.2807-4.3792 & 2.8627 & 44 & 29.04 & 0.78 & 1.39 & 0.56 & 23.734 & 0.001 \\ 
\midrule\\[-13pt]
\multicolumn{10}{l}{\sc Sub-mm:}\\
F225W & 2.2807-2.2979 & 2.2893 & 2 & $>$27.12 & \nodata & $(1.0)^{\ddagger}$ & 1.23 & 25.640 & 0.023 \\ [-4pt]
F275W & 2.4030-3.0750 & 2.6551 & 5 & 27.45 & 0.28 & 3.94 & 0.28 & 24.058 & 0.002 \\ [-4pt]
F336W & 3.1569-3.7098 & 3.5069 & 3 & 26.95 & 0.44 & 2.45 & 0.56 & 24.175 & 0.005 \\ [-4pt]
All LyC & 2.4030-3.7098 & 2.9745 & 8 & 29.94 & 0.41 & 2.66 & 0.37 & 24.442 & 0.002 \\ 
\midrule\\[-13pt]
\multicolumn{10}{l}{\sc Microwave detected:}\\
F225W & 2.2807-2.2979 & 2.2893 & 2 & $>$27.12 & \nodata & $(1.0)^{\ddagger}$ & 1.23 & 25.640 & 0.023 \\ [-4pt]
F275W & 2.4030-3.0750 & 2.6305 & 3 & 27.61 & 0.49 & 2.23 & 0.28 & 23.500 & 0.001 \\ [-4pt]
F336W & 3.9050-3.9050 & 3.9050 & 1 & $>$28.19 & \nodata & $(1.0)^{\ddagger}$ & 0.15 & 26.543 & 0.038 \\ [-4pt]
All LyC & 2.4030-3.9050 & 2.9491 & 4 & 30.17 & 0.44 & 2.45 & 0.27 & 24.045 & 0.002 \\ 
\midrule\\[-13pt]
\multicolumn{10}{l}{\sc Radio detected:}\\
F225W & 2.2807-2.2979 & 2.2893 & 2 & $>$27.12 & \nodata & $(1.0)^{\ddagger}$ & 1.23 & 25.640 & 0.023 \\ [-4pt]
F275W & 2.3860-3.0750 & 2.5986 & 12 & $>$28.61 & \nodata & $(1.0)^{\ddagger}$ & 0.24 & 22.838 & 0.000 \\ [-4pt]
F336W & 3.1569-3.6609 & 3.4712 & 4 & 27.62 & 0.54 & 2.00 & 0.51 & 24.437 & 0.005 \\ [-4pt]
All LyC & 2.3860-3.6609 & 2.8167 & 16 & $>$30.78 & \nodata & $(1.0)^{\ddagger}$ & 0.26 & 23.268 & 0.000 \\ 
\enddata
\tablecomments{\textbf{Table columns:} (1) Observed WFC3/UVIS filter; (2) Redshift range of galaxies included in LyC/UVC stacks. Overlapping ranges for the low- and high-$z$ ``All'' samples are due to splitting the F225W, F275W, and F336W samples by their median redshift before combining the three subsamples into either the high- or low-$z$ ``All'' sample; (3) Average redshift of all galaxies in each stack; (4) Number of galaxies with reliable spectroscopic redshifts included in each stack; (5) Observed total AB magnitude of LyC emission from stack (\SExtractor\ \texttt{MAG\_AUTO}) aperture matched to UVC, indicated by the blue ellipses in figs. X--Y; (6) 1$\sigma$ uncertainty in LyC AB-mag; (7) Measured S$/$N of the LyC stack flux (\boldmath$^{\dagger}$\unboldmath indicates a 1$\sigma$ upper limit); (8) Area (in arcsec$^2$) of the UVC aperture; (9) Observed total AB magnitude of the UVC stack; (10) 1$\sigma$ uncertainties of UVC AB-mag. Listed uncertainties do not include systematic filter zeropoint uncertainty, which are $<$\,2--4\%\ \citep{Windhorst2022,Obrien2023}.}
\end{deluxetable*} 

\begin{figure*}[ht!]\centerline{
\includegraphics[width=\textwidth]{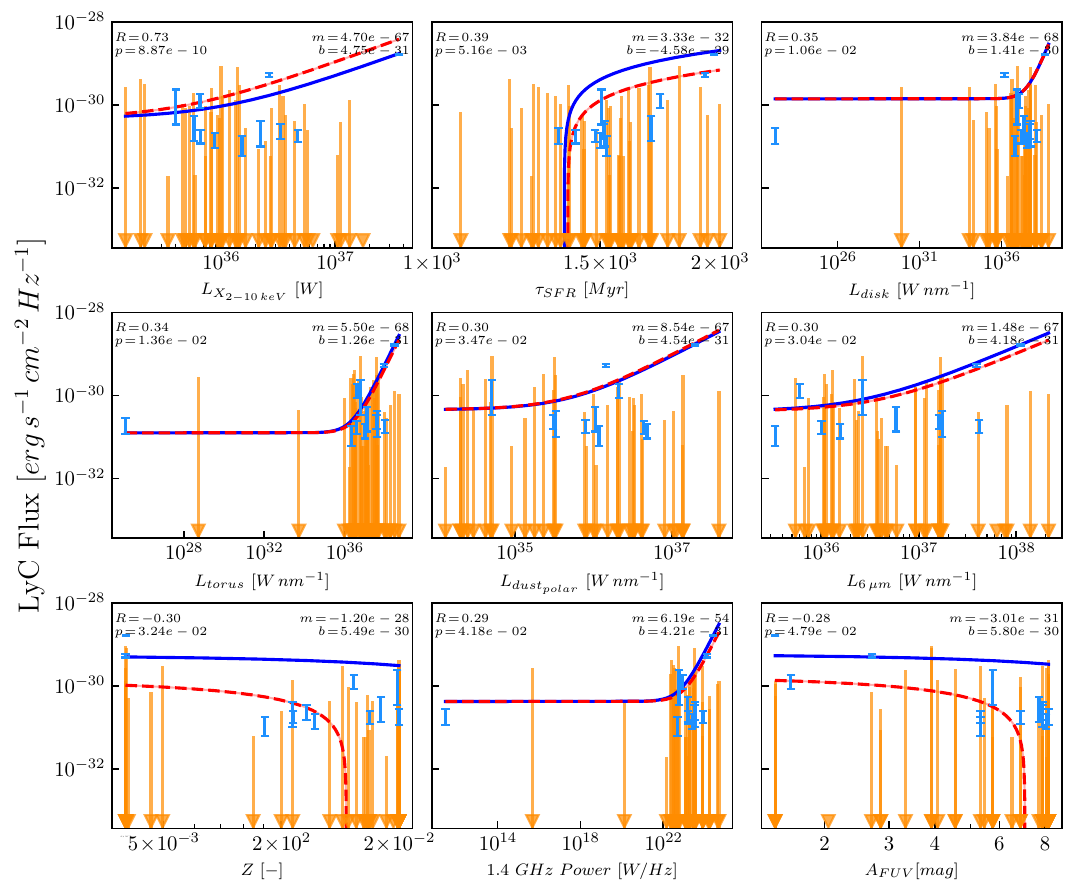}}
\caption{\label{corr}Lyman continuum flux vs. various best-fitting SED parameters with strong correlations for our AGN sample. Light-blue points indicate detected LyC while orange bars are 1$\sigma$ upperlimits to the LyC flux. The dark blue line is the Bayesian fit to the data with slope and intercept shown ($m$ and $b$, respectively), treating upper limits as censored and the red-dashed line is the same fit, excluding the brightest point with the largest weight. Despite this exclusion, the general trends still survive. All correlation coefficients ``$R$'' shown reject the null-hypothesis ($p<0.05$). The strongest correlating parameter is the X-ray luminosity from 2--10\,keV, indicating a strong X-ray to LyC emission connection for AGN. The SFH $\tau$ onset age parameter shows strong correlation, which implies higher LyC emission after more recent star-formation episodes. Several dust parameters (torus dust luminosity, polar dust luminosity, 6\,$\mu m$ luminosity, and A$_{FUV}$) show strong correlation with LyC flux, as well as 1.4 GHZ radio emission. Stellar metallicity shows a strong anti-correlation to LyC flux, which may signify a preference for higher LyC emission from metal poor galaxies. }
\end{figure*}

\vspace{-2\baselineskip}

\section{Results}
As seen in \S\ref{sample}, our AGN with LyC flux detected at $>$4$\sigma$ and those with LyC measured below 4$\sigma$ are most likely not distinct samples when considering the SED parameters we used for characterization. In order to ascertain some parameter dependence on the LyC flux, we computed the Pearson correlation coefficient between LyC flux, including upper limits, and several SED parameters, namely the $M_{AB}$, stellar mass, the ISM $A_V$, metallicity, the instantaneous star-formation rate, the ionization parameter $U$, the X-ray photon index, the AGN inclination, and the luminosities of the young, old, and young+old stellar populations, the AGN disk, the AGN dust torus, AGN polar dust, the disk+torus+polar dust, the ISM dust, and the X-ray emission. Since our LyC photometric data is composed of mostly upper limits, the LyC flux is most appropriate to correlate parameters to, since scaling the UV skybackground to high redshift luminosity would only increase the uncertainty in the data. Furthermore, flux correlations can inform future searches for AGN LyC leaker \emph{detections} in imaging. 

We found the parameter with the strongest correlation was best fitting AGN SED 2--10\,keV X-ray luminosity, with $r$\,$\simeq$\,0.73 and $p$\,$\simeq$\,8.86\,$\times$\,$10^{-10}$. The next strongest correlation was main stellar population's star-formation onset age $\tau$ ($r,\ p\,\simeq\,(0.39,\ 5.16\,\times\,10^{-3})$), assuming the delayed ``burst-quench'' model of \citet{Ciesla2017}. The remaining significant positive correlation coefficients in decreasing strengths are the AGN disk luminosity ($r,\ p\,\simeq\,(0.35,\ 0.011)$), AGN torus luminosity ($r,\ p\,\simeq\,(0.34,\ 0.014)$), the AGN monochromatic 6\,$\mu m$ luminosity ($r,\ p\,\simeq\,(0.30,\ 0.030)$), then the AGN 1.4 GHz power ($r,\ p\,\simeq\,(0.29,\ 0.042)$). We find a weak correlation to stellar luminosity, but this was not statistically significant ($p$\,$\simeq$\,0.19), which was stronger for the older stellar population ($p$\,$\simeq$\,0.10) compared to the younger population. We also find significant anti-correlation to the metallicity of the stellar component SED with $r,\ p\,\simeq\,(-0.30,\ 0.032)$ and $A_{FUV}$ ($r,\ p\,\simeq\,(-0.28,\ 0.048)$). We find no correlation to the AGN fraction parameter or X-ray photon index. 

In fig.~\ref{corr}, we show LyC flux vs. these parameters along with a Bayesian fit to the data. We used normal distributions to model the LyC flux probability for data with S/N\,$>$\,1$\sigma$, and treat the LyC flux with S/N\,$<$\,1$\sigma$ as censored data, modelling these data using arbitrary probability distributions while constraining the log-cumulative distribution function to the 1$\sigma$ uncertainty of the observed LyC flux. Since the points with highest SNR have the most weight in these fits, we also fit a line to the data excluding the point with the highest flux (plotted as a dashed-red curve). As shown, this does not have significant impacts on the general direction of these lines, though the curves for metallicity and $A_{FUV}$ show the most change. Of the eight curves, five are negligibly different, and the curves for $L_{X_{2-10\,keV}}$ and $\tau_{SFR}$ are within the uncertainty ranges of the data. The p-values are generally higher without this bright point, which indicates that these correlations would significantly improve with more LyC-detected AGN included in the analysis.

\section{Discussion}
In general, our results indicate that higher LyC emission can be traced to intrinsically higher LyC production via the AGN accretion disk and host-galaxy stellar sources. We compared subsamples of our AGN split into redshift bins and detected regions of the AGN continuum regions and find no significant distinctions. Our correlation tests show that high-LyC emitting AGN can be found from various markers that might indicate high ionizing photon production, i.e., X-ray and IR/dust emission brightness. X-ray emission from AGN is created as a result of Comptonization of optical/UV photons produced by the accretion disk, which up-scatter off of a corona of hot electrons in close proximity to the SMBH into the X-ray regime \citep{Haardt1991}. A significant fraction of these UV photons will be ionizing, therefore the LyC that is not processed into X-rays can escape the host-galaxy and ionize the IGM. Thus it is not surprising that X-ray luminosity correlates significantly with observed LyC flux. 

Some of the X-ray emission is also reprocessed by the dust torus, the broad-line region and the accretion disk. The toroidal and polar dust will become heated by the UV/optical light produced by the accretion disk, giving rise to extinction and luminous IR features in the SED. It is therefore reasonable to predict that brighter LyC production would result in brighter dust emission features, which is what we observed from correlating these parameters. The LyC that is not absorbed or reprocessed by dust will also have some probability of escaping the host-galaxy, and more flux will be observed if more LyC is produced by the disk. The detected LyC emission may rely on the line of sight or angle at which the AGN is observed.

Stellar sources in the host-galaxy will also contribute to the observed LyC since the accretion disk would mostly saturate LyC absorbers in the ISM and near the SMBH. This is true if the stellar sources can produce LyC in significant amounts, which are typically A, B, and O-type stars. The most massive O-type stars will produce the most LyC, though they typically only live for $\lesssim$\,10\,Myr before going supernova, $\lesssim$\,200\,Myr for B-types, and several hundred Myr for A-types. Therefore, one would expect that recent star-formation episodes producing newly formed massive stars would increase the observed LyC emitted by galaxies, which is corroborated by our correlation results. Furthermore, metal poor stars should produce more LyC than more metal rich stars since the photospheres would absorb less of the high energy ionizing light, which we also observe in our correlation analysis. 

We therefore infer that AGN with detectable LyC flux would be those with bright X-ray or thermal dust emission, with additional flux coming from the host-galaxy stellar population produced within a few Myrs of a star formation episode that produced lower metallicity stars. Given the generally fainter luminosities of our AGN sample, a natural follow up study would be to include brighter AGN with more robust LyC detections outside the UVCANDELS fields, such as those in \citet{Grazian2018} and \citet{Romano2019}, which could potentially result in more accurate correlation fits.

\section{Conclusions}
We measured the escaping LyC flux from 51 AGN with high quality spectroscopic redshifts found in the UVCANDELS, ERS, HDUV, and UVUDF data after an exhaustive search of Vizier AGN databases. We found 12 AGN with LyC emission where the detection reached a SNR\,$>$\,4$\sigma$. We used best-fitting CIGALE SEDs to model all of our AGN intrinsic flux and physical parameters using the highest SNR photometric data retrieved from NED database queries, ranging from X-ray to radio. We used these models, along with the IGM MC code of \citet{Inoue2014}, to calculate the \fesc\ of the 12 AGN with LyC detections. We correlated the SED parameters with the LyC flux of our AGN and fit lines to these correlations using a Bayesian regression, treating non-detections as censored data. We also stacked LyC cutout images of subsamples of AGN based on detected regions in the electromagnetic spectrum using the NED photometry, as well as subsamples of AGN based on the proximity of their LyC filter cutoff to their observed-frame Lyman-limit. Our results are as follows:

\noindent (1) The CIGALE SED fits of our 51 AGN indicate a weak, dusty, and mildly star-forming AGN population. From KS-tests, the 12 LyC-detected AGN are not distinguishable from the non-LyC-detected AGN in SED parameter histograms. The LyC-detected AGN \fesc\ values show a wide range, from 1\%\,$\lesssim$\,\fesc\,$\lesssim$\,100\%, or more (indicative of observed LyC flux outshining model predictions), with an inverse-variance weighted-average \fesc\,$\simeq$\,18\%. 

\noindent (2) We find no significant correlations of the LyC emission for the stacked subsamples based on their detections in the various regions of the AGN continuum. The brightest substack in LyC was the sub-mm stack in F336W (3.15\,$<$\,$z$\,$<$\,3.71), although this stack had three AGN and likely has a detection bias. The full stack of all 51 AGN was the brightest stack compared to stacks of all AGN detected in some region of the AGN continuum. The stack of all galaxies in the F336W sample (3.08\,$<$\,$z$\,$<$\,4.88 \zmean\,$\sim$\,3.5924) containing 16 AGN was the brightest substack with \mAB\,=\,27.66$\pm$0.56, outshining the two lower redshift bins at 2.28\,$<$\,$z$\,$<$\,2.30 and 2.38\,$<$\,$z$\,$<$\,3.08. After splitting each of these three samples by their median redshift into higher and lower redshift bins, then stacking these subsamples, we find that the lower-$z$ substacks show brighter LyC flux. 

\noindent (3) On an individual basis, we find significant correlation between the observed LyC flux and the 2--10\,keV X-ray luminosity, the main stellar population's star-formation onset age, AGN disk luminosity, AGN torus luminosity, the AGN monochromatic 6\,$\mu m$ luminosity, and the AGN 1.4 GHz power. We also find significant anti-correlation of the measured LyC flux with stellar metallicity and $A_{FUV}$. These findings indicate a connection to the emitted LyC flux and X-ray and thermal dust emission, and indicate that more recent star-formation episodes produce higher LyC emission, where LyC emission also favors metal poor galaxies. 

Although these correlations may not be exactly linear in nature, they do suggest a monotonically increasing relationship to observed LyC flux. The result of only finding 12 detections out of 51 AGN, with no evidence for redshift dependence, shows that the physical environments of the AGN and host-galaxy must play an intimate role in the escape of LyC. Our sample of AGN is still small, and only covers a redshift range from $z$\,$\sim$\,2.3--4.9. However, we are beginning to ascertain the physical origins for LyC escape in AGN with this small sample. This analysis should be expanded upon, both with more HST WFC3/UVIS data and to lower redshift using data from higher energy observatories. Expanding the sample to higher redshifts may be challenging given the increasing opacity of the IGM. Since AGN show essentially no evolution in their SED over cosmic time, the correlation of their physical parameters to their LyC emission could reveal the true impact of AGN on reionization. Their inferred LyC flux at the most relevant redshifts at $z$\,$\geq$\,6 could ideally be obtained simply by finding where these high-$z$ AGN lie on some correlated curves of LyC flux vs. a measurable physical parameter. 

\begin{acknowledgments}
We thank A. Inoue for providing the essential IGM attenuation montecarlo code used for \fesc\ calculations. This work is based on observations with the NASA/ESA Hubble Space Telescope obtained at the Space Telescope Science Institute, which is operated by the Association of Universities for Research in Astronomy, Incorporated, under NASA contract NAS5-26555. Support for Program number HST-GO-15647 was provided through a grant from the STScI under NASA contract NAS5-26555. RAW acknowledges support from NASA JWST Interdisciplinary Scientist grants NAG5-12460, NNX14AN10G and 80NSSC18K0200 from GSFC. XW acknowledges work carried out, in part, by IPAC at the California Institute of Technology, and was sponsored by the National Aeronautics and Space Administration. XW is supported by the Fundamental Research Funds for the Central Universities, and the CAS Project for Young Scientists in Basic Research, Grant No. YSBR-062. This research has made use of the NASA/IPAC Extragalactic Database (NED), which is funded by the National Aeronautics and Space Administration and operated by the California Institute of Technology.
\end{acknowledgments}

\facilities{\HST(ACS, WFC3)}

\software{astropy \citep{astropy2013,astropy2018,astropy2022}, 
          CIGALE \citep{Boquien2019},
          SExtractor \citep{Bertin1996}
          }
\bibliography{ms}

\appendix

\movetabledown=0.6in
\begin{longrotatetable}
\begin{deluxetable*}{lllccccccccr}
\tablecaption{AGN Photometric Properties\label{phottable}}
\tablehead{
\colhead{ID} & \colhead{RA} & \colhead{Dec} & \colhead{$z$} & \colhead{X-ray} & \colhead{IR} & \colhead{MIR} & \colhead{FIR} & \colhead{SMM} & \colhead{Microwave} & \colhead{Radio}\\[-5pt]
\colhead{(1)} & \colhead{(2)} & \colhead{(3)} & \colhead{(4)} & \colhead{(5)} & \colhead{(6)} & \colhead{(7)} & \colhead{(8)} & \colhead{(9)} & \colhead{(10)} & \colhead{(11)}}
\startdata
EIS J033201.59-274327.2 & 53.006594 & --27.724166 & 2.7212 & 31.29$\pm$0.03$^5$ & 22.53$\pm$0.01$^3$ & 21.93$\pm$0.01$^1$ & 18.82$\pm$0.03$^1$ & \nodata & \nodata & \nodata\\[-4pt]
CANDELS J033202.12-274327.6$^{\dagger}$ & 53.008878 & -27.724353 & 2.726 & \nodata & 24.00$\pm$0.03$^1$ & 23.38$\pm$0.02$^1$ & \nodata & \nodata & \nodata & \nodata\\[-4pt]
CANDELS J033203.03-274450.0 & 53.01265 & --27.747243 & 2.573 & 33.54$\pm$0.08$^3$ & 22.64$\pm$0.01$^3$ & 21.75$\pm$0.01$^1$ & 20.61$\pm$0.40$^1$ & \nodata & \nodata & \nodata\\[-4pt]
CANDELS J033203.24-274518.8$^{\dagger}$ & 53.013517 & -27.755213 & 3.2171 & \nodata & 24.10$\pm$0.03$^3$ & 24.61$\pm$0.05$^1$ & 20.21$\pm$0.10$^1$ & \nodata & \nodata & \nodata\\[-4pt]
EIS U21$^{\dagger}$ & 53.020574 & --27.742151 & 3.4739 & \nodata & 23.21$\pm$0.01$^3$ & 22.61$\pm$0.01$^1$ & 20.08$\pm$0.08$^1$ & \nodata & \nodata & \nodata\\[-4pt]
COMBO-17 35278 & 53.033345 & --27.782562 & 2.6123 & 31.93$\pm$0.04$^5$ & 23.46$\pm$0.02$^3$ & 21.53$\pm$0.00$^1$ & 18.46$\pm$0.02$^1$ & \nodata & \nodata & \nodata\\[-4pt]
LQAC 053-027 007$^{\dagger}$ & 53.034447 & --27.698209 & 2.47 & 35.08$\pm$0.22$^4$ & 24.02$\pm$0.02$^3$ & 22.54$\pm$0.01$^1$ & 20.60$\pm$0.21$^1$ & \nodata & \nodata & \nodata\\[-4pt]
WISEA J033209.44-274807.3 & 53.039367 & --27.801888 & 2.81 & 31.62$\pm$0.03$^5$ & 19.75$\pm$0.00$^1$ & 17.91$\pm$0.00$^3$ & 16.57$\pm$0.00$^1$ & \nodata & \nodata & \nodata\\[-4pt]
CANDELS J033217.41-274439.9 & 53.072567 & --27.744446 & 2.6503 & \nodata & 24.74$\pm$0.03$^1$ & \nodata & \nodata & \nodata & \nodata & \nodata\\[-4pt]
LQAC 053-027 027$^{\dagger}$ & 53.078474 & --27.859865 & 3.6609 & 33.64$\pm$0.09$^5$ & 24.53$\pm$0.02$^1$ & 21.16$\pm$0.00$^1$ & 18.90$\pm$0.05$^1$ & \nodata & \nodata & 14.69$\pm$0.25$^2$\\[-4pt]
CANDELS J033219.80-274518.9 & 53.08252 & --27.755248 & 2.926 & 33.16$\pm$0.08$^5$ & 24.03$\pm$0.09$^3$ & 22.11$\pm$0.01$^1$ & 19.56$\pm$0.04$^1$ & \nodata & \nodata & \nodata\\[-4pt]
CANDELS J033222.17-274936.5 & 53.092455 & --27.826868 & 2.54 & 36.49$\pm$0.27$^4$ & 25.02$\pm$0.04$^1$ & 21.01$\pm$0.00$^1$ & 19.52$\pm$0.07$^1$ & \nodata & \nodata & 19.74$\pm$0.15$^3$\\[-4pt]
CANDELS J033222.44-274543.8 & 53.093546 & --27.762193 & 2.2807 & 35.40$\pm$0.16$^4$ & 24.87$\pm$0.05$^3$ & 22.45$\pm$0.01$^1$ & 19.28$\pm$0.06$^1$ & \nodata & \nodata & \nodata\\[-4pt]
CANDELS J033222.53-274804.3 & 53.093908 & --27.801189 & 2.417 & 35.33$\pm$0.10$^4$ & 24.04$\pm$0.04$^3$ & 21.54$\pm$0.00$^2$ & 22.63$\pm$1.70$^1$ & \nodata & \nodata & 19.55$\pm$0.12$^1$\\[-4pt]
GMASS 2443 & 53.100804 & --27.715958 & 2.2979 & 33.97$\pm$0.10$^4$ & 22.43$\pm$0.01$^3$ & 21.33$\pm$0.00$^1$ & 19.62$\pm$0.05$^1$ & \nodata & \nodata & \nodata\\[-4pt]
CANDELS J033225.81-275120.4 & 53.107413 & --27.855551 & 2.879 & 35.26$\pm$0.14$^4$ & 24.19$\pm$0.05$^3$ & 22.41$\pm$0.01$^1$ & 20.91$\pm$0.15$^1$ & \nodata & \nodata & \nodata\\[-4pt]
CANDELS J033225.94-274514.3 & 53.108108 & --27.753984 & 2.728 & 33.28$\pm$0.06$^5$ & 23.71$\pm$0.02$^3$ & 22.15$\pm$0.01$^1$ & 20.99$\pm$0.46$^1$ & \nodata & \nodata & \nodata\\[-4pt]
CANDELS J033226.77-274603.9$^{\dagger}$ & 53.111576 & -27.767792 & 3.3313 & 34.26$\pm$0.09$^2$ & 26.44$\pm$0.04$^3$ & 22.28$\pm$0.01$^1$ & 20.54$\pm$0.20$^1$ & \nodata & \nodata & \nodata\\[-4pt]
CANDELS J033228.27-274403.4 & 53.117881 & --27.73433 & 3.256 & 35.70$\pm$0.15$^4$ & 22.11$\pm$0.03$^3$ & 20.97$\pm$0.00$^1$ & 20.02$\pm$0.14$^1$ & \nodata & \nodata & \nodata\\[-4pt]
CANDELS J033228.77-274434.9 & 53.119858 & --27.743092 & 3.0804 & 36.48$\pm$0.25$^4$ & 25.22$\pm$0.11$^3$ & 21.43$\pm$0.01$^2$ & \nodata & \nodata & \nodata & \nodata\\[-4pt]
LQAC 053-027 039$^{\dagger}$ & 53.124379 & --27.851638 & 3.7098 & 32.70$\pm$0.06$^5$ & 23.53$\pm$0.02$^3$ & 21.28$\pm$0.00$^1$ & 18.81$\pm$0.02$^1$ & 15.71$\pm$0.67$^1$ & \nodata & 15.65$\pm$1.41$^2$\\[-4pt]
CANDELS J033232.16-274651.4 & 53.134022 & --27.780974 & 2.542 & 33.58$\pm$0.08$^5$ & 24.31$\pm$0.02$^3$ & 22.06$\pm$0.01$^1$ & 19.47$\pm$0.04$^1$ & \nodata & \nodata & \nodata\\[-4pt]
GMASS 2578 & 53.137581 & --27.700116 & 2.4481 & \nodata & 22.66$\pm$0.01$^3$ & 20.87$\pm$0.00$^1$ & 18.82$\pm$0.04$^1$ & \nodata & \nodata & \nodata\\[-4pt]
CANDELS J033233.24-274916.0 & 53.13851 & --27.821132 & 3.4836 & 37.22$\pm$0.25$^4$ & 26.00$\pm$0.02$^1$ & 23.89$\pm$0.02$^1$ & \nodata & \nodata & \nodata & \nodata\\[-4pt]
GMASS 2562 & 53.138699 & --27.700492 & 2.4495 & \nodata & 23.01$\pm$0.02$^3$ & 20.82$\pm$0.00$^1$ & 19.78$\pm$0.10$^1$ & \nodata & \nodata & \nodata\\[-4pt]
LQAC 053-027 045 & 53.148832 & --27.821104 & 2.576 & 34.63$\pm$0.12$^2$ & 22.22$\pm$0.01$^3$ & 20.97$\pm$0.00$^1$ & 17.16$\pm$0.01$^1$ & \nodata & \nodata & 19.65$\pm$0.14$^3$\\[-4pt]
GOODS-CDFS-MUSIC 15985 & 53.157456 & --27.709059 & 2.9768 & \nodata & 23.31$\pm$0.02$^3$ & 22.97$\pm$0.01$^1$ & 18.01$\pm$0.01$^1$ & \nodata & \nodata & \nodata\\[-4pt]
CANDELS J033238.15-274626.7 & 53.159 & --27.774097 & 3.7148 & \nodata & 28.57$\pm$0.05$^1$ & \nodata & \nodata & \nodata & \nodata & \nodata\\[-4pt]
GALEXASC J033238.92-274628.5 & 53.162258 & --27.774725 & 2.447 & \nodata & 20.55$\pm$0.00$^1$ & 20.77$\pm$0.00$^1$ & 18.90$\pm$0.03$^1$ & \nodata & \nodata & \nodata\\[-4pt]
GMASS 2467 & 53.162299 & --27.71213 & 4.3792 & 34.41$\pm$0.19$^2$ & 22.04$\pm$0.01$^3$ & 20.72$\pm$0.00$^1$ & 19.19$\pm$0.04$^1$ & \nodata & \nodata & \nodata\\[-4pt]
GMASS 0253 & 53.163215 & --27.808983 & 4.8759 & 35.10$\pm$0.19$^2$ & 23.91$\pm$0.00$^4$ & 22.20$\pm$0.01$^1$ & 22.68$\pm$1.64$^1$ & \nodata & \nodata & \nodata\\[-4pt]
GMASS 2043 & 53.174458 & --27.733302 & 2.576 & 32.82$\pm$0.06$^5$ & 23.82$\pm$0.02$^3$ & 22.28$\pm$0.01$^1$ & 21.06$\pm$0.51$^1$ & \nodata & \nodata & \nodata\\[-4pt]
CANDELS J033242.83-274702.5 & 53.178486 & --27.784035 & 3.166 & 32.61$\pm$0.05$^5$ & 23.80$\pm$0.00$^2$ & 23.09$\pm$0.01$^1$ & 13.62$\pm$0.32$^1$ & \nodata & \nodata & \nodata\\[-4pt]
HUDF-JD1 & 53.178696 & --27.802639 & 2.4245 & 33.28$\pm$0.28$^1$ & 24.75$\pm$0.02$^4$ & 21.80$\pm$0.01$^2$ & 19.88$\pm$0.05$^1$ & \nodata & \nodata & \nodata\\[-4pt]
IRAC J033244.00-274635.0 & 53.183359 & --27.77639 & 2.688 & 33.45$\pm$0.07$^5$ & 23.33$\pm$0.00$^3$ & 21.15$\pm$0.00$^2$ & 18.38$\pm$0.01$^1$ & \nodata & \nodata & \nodata\\[-4pt]
SDSS J123557.63+621024.4 & 188.990099 & 62.173431 & 3.075 & 33.52$\pm$0.04$^4$ & 22.03$\pm$0.00$^1$ & 21.38$\pm$0.00$^1$ & 18.73$\pm$0.04$^1$ & 13.81$\pm$0.11$^2$ & 15.53$\pm$0.42$^1$ & 20.36$\pm$0.13$^3$\\[-4pt]
LQAC 189+062 004$^{\dagger}$ & 189.027919 & 62.264065 & 2.4135 & 32.94$\pm$0.04$^4$ & 22.80$\pm$0.01$^1$ & 21.50$\pm$0.00$^1$ & 18.32$\pm$0.03$^1$ & 14.08$\pm$0.18$^2$ & 16.09$\pm$0.48$^1$ & 20.38$\pm$0.19$^3$\\[-4pt]
S-CANDELS J123611.92+621147.6 & 189.049878 & 62.196518 & 2.403 & \nodata & 22.76$\pm$0.02$^2$ & 20.96$\pm$0.00$^1$ & 18.22$\pm$0.06$^1$ & 14.03$\pm$0.33$^3$ & 16.00$\pm$0.35$^1$ & 21.48$\pm$0.30$^3$\\[-4pt]
WISEA J123623.00+621526.8$^{\dagger}$ & 189.095583 & 62.257405 & 2.592 & 30.21$\pm$0.02$^2$ & 20.25$\pm$0.00$^1$ & 19.80$\pm$0.00$^1$ & 17.22$\pm$0.01$^1$ & 16.17$\pm$1.72$^2$ & \nodata & 20.00$\pm$0.08$^3$\\[-4pt]
LQAC 189+062 014 & 189.122658 & 62.253609 & 3.6539 & 35.08$\pm$0.04$^8$ & 22.75$\pm$0.00$^1$ & 22.04$\pm$0.00$^1$ & 19.39$\pm$0.05$^1$ & 14.58$\pm$0.22$^2$ & \nodata & 15.50$\pm$0.18$^2$\\
LQAC 189+062 016$^{\dagger}$ & 189.139593 & 62.238358 & 3.413 & 33.93$\pm$0.04$^6$ & 25.01$\pm$0.03$^1$ & 22.67$\pm$0.00$^1$ & 20.69$\pm$0.28$^1$ & \nodata & \nodata & 18.09$\pm$0.15$^3$\\[-4pt]
LQAC 189+062 026 & 189.186232 & 62.258635 & 2.453 & \nodata & 24.44$\pm$0.02$^1$ & 22.95$\pm$0.00$^1$ & 20.41$\pm$0.16$^1$ & \nodata & \nodata & 23.01$\pm$1.18$^3$\\[-4pt]
S-CANDELS J123652.47+621751.1 & 189.218479 & 62.297495 & 3.905 & \nodata & 24.92$\pm$0.02$^1$ & 24.01$\pm$0.02$^1$ & 19.52$\pm$0.08$^1$ & \nodata & 16.26$\pm$0.66$^1$ & 16.86$\pm$1.50$^2$\\[-4pt]
S-CANDELS J123655.82+621201.1 & 189.232475 & 62.200212 & 2.737 & 35.28$\pm$0.04$^7$ & 22.71$\pm$0.02$^1$ & 20.86$\pm$0.00$^1$ & 18.08$\pm$0.01$^1$ & 13.92$\pm$0.19$^2$ & \nodata & 14.92$\pm$0.08$^2$\\[-4pt]
LQAC 189+062 039 & 189.27002 & 62.353268 & 2.487 & \nodata & 24.56$\pm$0.06$^1$ & 20.93$\pm$0.15$^3$ & 18.81$\pm$0.11$^1$ & \nodata & \nodata & \nodata\\[-4pt]
S-CANDELS J123706.20+621844.2 & 189.275646 & 62.312224 & 2.414 & \nodata & 23.45$\pm$0.01$^1$ & 22.91$\pm$0.01$^1$ & \nodata & \nodata & \nodata & \nodata\\[-4pt]
S-CANDELS J123714.30+621208.5$^{\dagger}$ & 189.309519 & 62.20236 & 3.1569 & 34.15$\pm$0.04$^6$ & 22.29$\pm$0.00$^1$ & 20.71$\pm$0.00$^1$ & 18.05$\pm$0.02$^1$ & 13.84$\pm$0.19$^2$ & \nodata & 20.21$\pm$0.19$^3$\\[-4pt]
S-CANDELS J123719.43+621113.9 & 189.3309 & 62.187193 & 2.457 & \nodata & 22.93$\pm$0.01$^1$ & 21.56$\pm$0.00$^1$ & 18.82$\pm$0.04$^1$ & \nodata & \nodata & 22.12$\pm$0.53$^3$\\[-4pt]
LQAC 189+062 052 & 189.332675 & 62.165328 & 2.647 & 32.67$\pm$0.04$^6$ & \nodata & 20.95$\pm$0.00$^2$ & 18.42$\pm$0.02$^1$ & 14.78$\pm$0.49$^2$ & \nodata & 21.86$\pm$0.41$^3$\\[-2pt]
S-CANDELS J123733.78+621345.2$^{\dagger}$ & 189.390877 & 62.22909 & 2.386 & \nodata & 22.24$\pm$0.01$^1$ & 20.73$\pm$0.00$^1$ & 18.03$\pm$0.01$^1$ & 15.64$\pm$1.71$^2$ & \nodata & 22.17$\pm$0.56$^3$\\[-4pt]
SDSS J123757.29+621627.3 & 189.488797 & 62.274298 & 2.9394 & 31.75$\pm$0.04$^4$ & 21.61$\pm$0.00$^1$ & 20.42$\pm$0.00$^1$ & 18.37$\pm$0.03$^1$ & \nodata & \nodata & 22.36$\pm$0.74$^3$\\[-2pt]
\enddata
\tablecomments{\\ Tabulated AB-magnitudes of the highest SNR band in the following spectral ranges:\\ X-ray: $^1$Chandra 0.2--0.5\,keV, $^2$Chandra 0.5--10\,keV, $^3$Chandra 0.5--1.2\,keV, $^4$Chandra 0.5--2\,keV, $^5$Chandra 0.5--7\,keV, $^6$Chandra 0.5--8\,keV, $^7$Chandra 2--8\,keV, $^8$Chandra 4--8\,keV\\ IR: $^1$CFHT/WIRCam $K$, $^2$HST/WFC3-IR F125W, $^3$HST/WFC3-IR F160W, $^4$HST/NICMOS F160W, $^5$VLT/ISAAC  $K_s$, $^6$VISTA/VIRCAM $J$\\ MIR: Spitzer/IRAC $^1$3.6\,$\mu$m, $^2$4.5\,$\mu$m, $^3$5.8\,$\mu$m\\ FIR: $^1$Spitzer/MIPS 24\,$\mu$m\\ SMM: $^1$LABOCA 870\,$\mu$m, $^2$SPIRE 250\,$\mu$m, $^3$SPIRE 350\,$\mu$m\\ Microwave: $^1$AzTEC 1.16\,mm\\ Radio: $^1$ATCA/CABB, $^2$VLA L-band.\\ $^{\dagger}$ indicates a LyC leaker.}
\end{deluxetable*}
\end{longrotatetable}
\clearpage

\end{document}